\documentclass[aps,prl,amsfonts,amssymb,amsmath,twocolumn,superscriptaddress,showpacs]{revtex4}

\usepackage{graphicx}
\usepackage{bm}

\newcommand{\bmath}[1]{\mbox{{\boldmath{{$#1$}}}}}

\begin{document}

\title{Abundant stable gauge field hair for black holes in anti-de Sitter space}

\author{J. E. Baxter}
\affiliation{Department of Applied Mathematics, The University of
Sheffield, Hicks Building, Hounsfield Road, Sheffield, S3 7RH,
United Kingdom.}
\author{Marc Helbling}
\affiliation{INSA de Rouen, Laboratoire de Math\'ematiques (LMI),
Place Emile Blondel BP 08, 76131 Mont Saint Aignan Cedex, France.}
\author{Elizabeth Winstanley} \email{E.Winstanley@sheffield.ac.uk}
\affiliation{Department of Applied Mathematics, The University of
Sheffield, Hicks Building, Hounsfield Road, Sheffield, S3 7RH,
United Kingdom.}

\date{\today }

\begin{abstract}
We present new hairy black hole solutions of ${\mathfrak {su}}(N)$ Einstein-Yang-Mills theory (EYM)
in asymptotically anti-de Sitter (adS) space.
These black holes are described by $N+1$ independent parameters, and have $N-1$ independent
gauge field degrees of freedom.
Solutions in which all gauge field functions have no zeros exist for all $N$, and for
sufficiently large (and negative) cosmological constant.
At least some of these solutions are shown to be stable under  classical, linear, spherically symmetric perturbations.
Therefore there is no upper bound on the amount of stable gauge field hair with which a black hole
in adS can be endowed.
\end{abstract}


\maketitle

The ``no-hair'' conjecture \cite{wheeler}
states that black hole equilibrium states possess extremely simple geometries, determined
completely by the mass, angular momentum and charge of the black hole.
While hairy black hole solutions of the Einstein equations have been discovered, particularly in
Einstein-Yang-Mills (EYM) theory and its variants (see \cite{volkov} for a review),
many of the plethora of new black hole solutions found in the literature are classically unstable.
Those hairy black holes which are stable (such as the ${\mathfrak {su}}(2)$
EYM black holes in anti-de Sitter space (adS) \cite{ew,bjoraker})
have, at least to date, been described by only a small number of
parameters additional to the mass, angular momentum and charge of the black hole.
This means that the ``spirit'' if not the ``letter'' of the no-hair conjecture is maintained.

In recent years there has been an explosion of interest in hairy black holes in adS, partly because at least
some of these configurations are stable, but also because of the importance of the adS/CFT correspondence
\cite{adsCFT} in string theory.
In particular, it has been suggested \cite{hertog} that there should be observables in the dual (deformed) CFT
which are sensitive to the presence of black hole hair (see also \cite{gauntlett} for an adS/CFT interpretation
of some stable seven-dimensional black holes with ${\mathfrak {so}}(5)$ gauge fields).
Our purpose in this letter is to present new stable, asymptotically adS,
hairy black hole solutions of ${\mathfrak {su}}(N)$ EYM  for sufficiently large $\left| \Lambda \right| $ which
are described by an unbounded number of parameters.
The existence of these solutions casts the status of the ``no-hair'' conjecture in a completely new light: equilibrium
black holes in adS are no longer simple objects, but rather require an infinite number of parameters
in order to fully determine their geometry.

%Hairy black hole solutions in Einstein-Yang-Mills (EYM) theory and its variants have been the subject of
%great interest for nearly twenty years (see \cite{volkov} for a review).
%However, many of the plethora of new black hole solutions found in the literature are classically unstable.
%Notable exceptions include the ${\mathfrak {su}}(2)$ EYM black holes in anti-de Sitter space (adS) \cite{ew,bjoraker}.
%Black hole solutions for which the
%gauge field function $\omega (r)$ has no zeros (nodes) exist for sufficiently large $\left| \Lambda \right| $.
%Furthermore, at least some of these nodeless
%solutions are stable under linear, spherically symmetric perturbations (this was subsequently extended
%to cover non-spherically symmetric perturbations in \cite{sarbach}).
%In this letter we outline the extension of these solutions to ${\mathfrak {su}}(N)$ EYM, finding, for any
%$N$ and sufficiently large $\left| \Lambda \right| $, black hole solutions with $N-1$ gauge field
%degrees of freedom which are stable under classical, linear, spherically symmetric, perturbations.

We consider static, spherically symmetric, four-dimensional black holes with metric
\begin{equation}
ds^{2} = - \mu S^{2} \, dt^{2} + \mu ^{-1} \, dr^{2} +
r^{2} \, d\theta ^{2} + r^{2} \sin ^{2} \theta \, d\phi ^{2} ,
\label{eq:metric}
\end{equation}
where the metric functions $\mu $ and $S$ depend on the radial co-ordinate $r$ only.
Here, and throughout this letter, the metric has signature $(-,+,+,+)$ and we use units in which $4\pi G=c=1$.
In the presence of a negative cosmological constant $\Lambda$, we write the metric function $\mu $ as
\begin{equation}
\mu (r) = 1 - \frac {2m(r)}{r} - \frac {\Lambda r^{2}}{3}.
\end{equation}
The most general, spherically symmetric, ansatz for the ${\mathfrak {su}}(N)$ gauge potential has been given
in~\cite{kunzle}.
Here, we assume that the gauge potential is purely magnetic and has the gauge-fixed form:
\begin{equation}
{\cal {A}} = \frac {1}{2} \left( C - C^{H} \right) \, d\theta - \frac {i}{2} \left[ \left(
C + C^{H} \right) \sin \theta + D \cos \theta \right] \, d\phi ,
\label{eq:gaugepot}
\end{equation}
where $C$ is an $(N \times N)$ upper-triangular matrix with non-zero entries immediately above the diagonal:
\begin{equation}
C_{j,j+1}=\omega_j (r) ,
\end{equation}
for $j=1,\ldots,N-1$,
with $C^{H}$ the Hermitian conjugate of $C$, and $D$ is a constant diagonal matrix:
\begin{equation}
D=\mbox{Diag}\left(N-1,N-3,\ldots,-N+3,-N+1\right) .
\label{eq:matrixD}
\end{equation}
The $(N-1)$ Yang-Mills equations take the form
\begin{equation}
r^2\mu\omega''_{j}+\left(2m-2r^3 p_{\theta}-\frac{2\Lambda
r^3}{3}\right)\omega'_{j}+W_j\omega_j=0
\label{eq:YMe}
\end{equation}
for $j=1,\ldots,N-1$, where a prime $'$ denotes $d/dr $, and
\begin{eqnarray}
p_{\theta}&=&
\frac{1}{4r^4}\sum^N_{j=1}\left[\left(\omega^2_j-\omega^2_{j-1}-N-1+2j\right)^2\right],\\
W_j&=&
1-\omega^2_j+\frac{1}{2}\left(\omega^2_{j-1}+\omega^2_{j+1}\right),
\end{eqnarray}
with $\omega_0=\omega_N=0$.
The Einstein equations take the form
\begin{equation}
m' =
\mu G+r^2p_{\theta},
\qquad
\frac{S'}{S}=\frac{2G}{r},
\label{eq:Ee}
\end{equation}
where
\begin{equation}
G=\sum^{N-1}_{j=1}\omega_j'^2.
\end{equation}
The field equations (\ref{eq:YMe},\ref{eq:Ee}) have the following trivial solutions.
Setting $\omega _{j}(r) \equiv \pm {\sqrt {j(N-j)}}$ for all $j$ gives the Schwarzschild-adS black hole
with $m(r) =M= {\mbox {constant}}$ (which can be set to zero to give pure adS space).
Setting $\omega _{j}(r) \equiv 0 $ for all $j$ gives the Reissner-Nordstr\"om-adS black hole with magnetic charge.
%and metric function
%\begin{equation}
%\mu (r) = 1 - \frac {2M}{r} + \frac {1}{r^{2}} - \frac {\Lambda r^{2}}{3}.
%\end{equation}
There is an additional special class of solutions, given by setting
\begin{equation}
\omega _{j}(r) =\pm  {\sqrt {j(N-j)}} \, \omega (r) \qquad \forall j=1,\ldots ,N-1.
\label{eq:embeddedsu2}
\end{equation}
In this case, it is possible to show, using a rescaling method along the lines of that in \cite{kunzle1},
that the field variables $\omega (r)$, $m(r)$ and $S(r)$ satisfy the ${\mathfrak {su}}(2)$ EYM field equations
with a negative cosmological constant.
Furthermore, the boundary conditions (as discussed below) are also preserved.
Therefore any ${\mathfrak {su}}(2)$, asymptotically adS, EYM black hole solution can be embedded into
${\mathfrak {su}}(N)$ EYM to give another asymptotically adS black hole.

In this letter we study black hole solutions of the field equations (\ref{eq:YMe},\ref{eq:Ee}), returning to
soliton solutions elsewhere \cite{suNsolutions}.
We assume there is a regular, non-extremal, black hole event horizon at $r=r_{h}$.
The field variables $\omega _{j}(r)$, $m(r)$ and $S(r)$ will have regular Taylor series expansions about $r=r_{h}$.
%\begin{eqnarray}
%m(r) & = & m( r_{h} ) + m'( r_{h} ) \left( r - r_{h} \right)
%+ O \left( r- r_{h} \right) ^{2} ;
%\nonumber \\
%\omega _{j} (r) & = & \omega _{j}( r_{h} ) + O \left( r -r_{h} \right) ;
%\nonumber \\
%S(r) & = & S( r_{h} ) + O\left( r - r_{h} \right) ;
%\label{eq:horizon}
%\end{eqnarray}
%where
%\begin{equation}
%2m( r_{h} ) = r_{h} - \frac {\Lambda r_{h}^{3}}{3}.
%\end{equation}
These expansions are determined by the $N+1$ quantities $\omega _{j}(r_{h})$, $r_{h}$, $S(r_{h})$ for fixed
cosmological constant $\Lambda $.
Since the field equations (\ref{eq:YMe},\ref{eq:Ee}) are invariant under the transformation
$\omega _{j}(r) \rightarrow - \omega _{j}(r)$ (for any $j$ independently), we may consider
$\omega _{j}( r_{h} ) >0$ without loss of generality.
For the event horizon to be non-extremal, it must be the case that
\begin{equation}
2m'( r_{h} ) = 2r_{h}^{2} p_{\theta} ( r_{h} ) < 1- \Lambda r_{h}^{2},
\label{eq:constraint}
\end{equation}
which constrains the possible values of the gauge field functions $\omega _{j} ( r_{h} ) $
at the event horizon.
At infinity, the boundary conditions are considerably less stringent than in the asymptotically flat case.
In order for the metric (\ref{eq:metric}) to be asymptotically adS,
we simply require that the field variables $\omega _{j}(r)$, $m(r)$ and $S(r)$ converge to constant values as
$r\rightarrow \infty $, and have regular Taylor series expansions in $r^{-1}$ near infinity.
%\begin{eqnarray}
%m(r)  & = &  M + O \left( r^{-1} \right) ;
%\qquad
%S(r) = 1 + O\left( r^{-1} \right) ;
%\nonumber \\
%\omega _{j}(r) & = & \omega _{j,\infty } + O \left( r^{-1} \right) .
%\label{eq:infinity}
%\end{eqnarray}
Since $\Lambda <0$, there is no cosmological horizon.

The field equations (\ref{eq:YMe},\ref{eq:Ee}) are integrated numerically using standard `shooting' techniques
\cite{NR}.
The equation for $S(r)$ decouples from the other Einstein equation and the Yang-Mills equations so can be integrated
separately if required.
%For ${\mathfrak {su}}({N})$ solutions, we therefore have $N$ ordinary differential equations to integrate
%($N-1$ Yang-Mills equations and one Einstein equation).
We start integrating just outside the event horizon, using as our shooting parameters
the $N$ variables $\omega _{j}(r_{h})$ and $r_{h}$,
subject to the weak constraint (\ref{eq:constraint}).
The field equations are then integrated outwards in the radial co-ordinate $r$ until either the field variables start to
diverge or they have converged to the asymptotic form at infinity.

\begin{figure}[h]
\begin{center}
\includegraphics[width=6.5cm,angle=270]{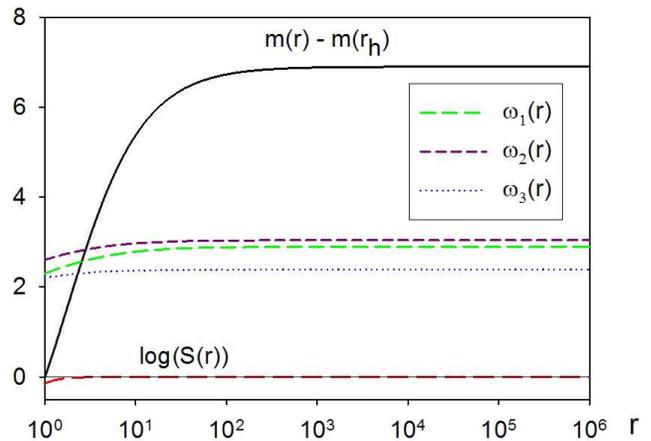}
\end{center}
\caption{A typical black hole solution of ${\mathfrak {su}}(4)$ EYM in which all the gauge
field functions $\omega _{j}(r)$ are nodeless.
For this solution, $\Lambda = -10$ and $r_{h}=1$.
The values of the gauge field functions on the event horizon are: $\omega _{1}( r_{h} ) = 2.3$,
$\omega _{2}( r_{h} ) = 2.6$ and $\omega _{3} ( r_{h} ) = 2.2$. }
\label{fig:example}
\end{figure}

As in the ${\mathfrak {su}}(2)$ case \cite{ew},
we find black hole solutions in open subsets of the $N$-dimensional parameter space
$\left( \omega _{j}(r_{h}),r_{h}\right) $ for fixed $\Lambda $.
For sufficiently large $\left| \Lambda \right| $ (where how large ``sufficiently large'' is depends on the radius
of the event horizon $r_{h}$), we find that the gauge field functions $\omega _{j}(r)$
all have no zeros.
In figure \ref{fig:example} we show a typical nodeless solution, for ${\mathfrak {su}}(4)$ EYM.
It can be seen that the metric functions $m(r)$ and $S(r)$ have very similar behaviour to the ${\mathfrak {su}}(2)$
case, and that, since $\left| \Lambda \right| $ is so large,
 the gauge field functions do not vary significantly from their values at the event horizon.

\begin{figure}[h]
\begin{center}
\includegraphics[width=6.5cm,angle=270]{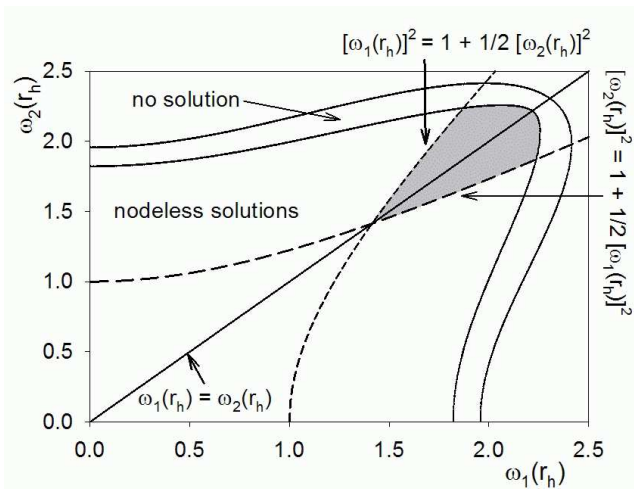}
\end{center}
\caption{Phase space of black hole solutions in ${\mathfrak {su}}(3)$ EYM with $\Lambda = -10$ and $r_{h}=1$.
The shaded region shows where solutions exist which satisfy the inequalities (\ref{eq:stabineq}) at the
event horizon.}
\label{fig:su3stab}
\end{figure}

The phase space of black hole solutions in the ${\mathfrak {su}}(3)$ case, with $\Lambda = -10$ and $r_{h}=1$ is shown
in figure \ref{fig:su3stab}, and is typical of the phase space for large values of $\left| \Lambda \right| $.
%\begin{figure}[h]
%\begin{center}
%\includegraphics[width=6.5cm,angle=270]{su3_phase.eps}
%\end{center}
%\caption{Phase space of black hole solutions in ${\mathfrak {su}}(3)$ EYM with $\Lambda = -10$ and $r_{h}$=1.}
%\label{fig:su3phase}
%\end{figure}
In figure \ref{fig:su3stab} we have examined, for $\Lambda = -10$ and $r_{h}=1$, all values of the
$\omega _{1}( r_{h} ) $ and $\omega _{2} ( r_{h} ) $ which satisfy the constraint
(\ref{eq:constraint}).
The inequality in (\ref{eq:constraint}) is saturated on the outer-most curve in figure \ref{fig:su3stab}.
It can be seen from figure \ref{fig:su3stab} that not all values of
$\left( \omega _{1}( r_{h} ), \omega _{2} ( r_{h} ) \right) $ give
black hole solutions; those values for which no regular black hole solution satisfying the boundary conditions
at infinity
%(\ref{eq:infinity})
could be found lie in the narrow band on the outside of the plot.
The region between this narrow band and the co-ordinate axes contains
black hole solutions in which both gauge field functions $\omega _{1}(r)$
and $\omega _{2}(r)$ have no zeros.
We have also plotted in figure \ref{fig:su3stab} the line
$\omega _{1}( r_{h} ) = \omega _{2} ( r_{h} ) $, on which lie embedded ${\mathfrak {su}}(2)$
solutions given by (\ref{eq:embeddedsu2}).
The significance of the shaded region in figure \ref{fig:su3stab} will be described shortly.
More detailed properties of the phase space of black hole solutions
%particularly for smaller values of the cosmological constant,
will be discussed elsewhere \cite{suNsolutions}.

In \cite{ew}, the existence of black hole solutions for which the gauge function $\omega (r)$ had no zeros was
proven analytically in the ${\mathfrak {su}}(2)$ case.
Since ${\mathfrak {su}}(2)$ solutions can be embedded as ${\mathfrak {su}}(N)$ solutions via (\ref{eq:embeddedsu2}),
we have automatically an analytic proof of the existence of nodeless ${\mathfrak {su}}(N)$ EYM black holes in adS.
However, these embedded solutions are `trivial' in the sense that they are described by just three parameters:
$r_{h}$, $\Lambda $ and $\omega ( r_{h} ) $.
An important question is whether the existence of `non-trivial' (that is, genuinely ${\mathfrak {su}}(N)$)
solutions in which all the gauge field functions $\omega _{j}(r)$ have no zeros can be proven analytically.
The answer to this question is affirmative, and involves a generalization to ${\mathfrak {su}}(N)$ of the
continuity-type argument used in \cite{ew}.
The details are lengthy and will be presented elsewhere.
However, the main thrust of the argument can be simply stated.
%For any $N$, we have embedded ${\mathfrak {su}}(2)$ solutions which have no nodes.
We firstly prove (generalizing the analysis of \cite{kunzle1} to include $\Lambda $)
that the field equations (\ref{eq:YMe},\ref{eq:Ee}) and initial conditions at the event horizon
%(\ref{eq:horizon})
possess, locally in a neighborhood of the horizon, solutions which are analytic in
$r$, $r_{h}$, $\Lambda $ and the parameters $\omega _{j} ( r_{h} ) $.
This enables us to prove that, in a sufficiently small neighborhood of any embedded ${\mathfrak {su}}(2)$
solution in which $\omega (r)$ has no nodes, there exists (at least in a neighborhood of the
event horizon) an ${\mathfrak {su}}(N)$ solution in which all the $\omega _{j}(r)$ have
no nodes.
The key part of the proof lies in then showing that these ${\mathfrak {su}}(N)$ solutions
can be extended out to $r\rightarrow \infty $ and that they
satisfy the boundary conditions
%(\ref{eq:infinity})
at infinity.
This gives genuinely ${\mathfrak {su}}(N)$ black hole solutions in which all the gauge
field functions have no zeros, and which are characterized by the $N+1$ parameters
$r_{h}$, $\Lambda $ and $\omega _{j} ( r_{h} ) $.
%In other words, for any $N$, we can prove analytically the existence of black hole solutions of the
%${\mathfrak {su}}(N)$ EYM equations which are described by the $N+1$ independent parameters
%$r_{h}$, $\Lambda $ and $\omega _{j} ( r_{h} ) $, for which all the gauge field functions
%have no zeros.

The other outstanding question is whether these new black holes,
with potentially unbounded amounts of gauge field hair, are stable.
We consider linear, spherically symmetric perturbations only for simplicity.
%The analysis of \cite{sarbach} in the ${\mathfrak {su}}(2)$ case
%revealed that, for sufficiently large $\left| \Lambda \right| $, stability under spherically symmetric perturbations
%continued to hold also for non-spherically symmetric perturbations, and one might hope that a similar result will
%hold in the more complex ${\mathfrak {su}}(N)$ case.
%However, we leave this for future work.
Even for spherically symmetric perturbations, the analysis is highly involved in the ${\mathfrak {su}}(N)$ case
and the details will be presented elsewhere.
Here we briefly outline just the key features.

Firstly we consider spherically symmetric perturbations of the gauge potential (\ref{eq:gaugepot}), fixing
the gauge so that the perturbed potential is purely magnetic and has the form \cite{kunzle}
\begin{eqnarray}
A & = &  {\cal {B}}  \, dr
+ \frac {1}{2} \left( C - C^{H} \right) d\theta
\nonumber \\ & &
- \frac {i}{2} \left[ \left( C + C^{H} \right) \sin \theta + D \cos \theta \right] d\phi .
\label{eq:pertpot}
\end{eqnarray}
Here, the matrices ${\cal {B}}$ and $C$ depend on both $t$ and $r$, and
matrix $D$ is still constant and given by (\ref{eq:matrixD}).
The matrix ${\cal {B}}(t,r)$ is traceless, diagonal and has purely imaginary entries.
The only non-zero entries of the matrix $C(t,r)$ are:
\begin{equation}
C_{j,j+1} (t,r) = \omega _{j} (t,r) \exp \left( i\gamma _{j}(t,r) \right).
\end{equation}
As usual, the metric retains the form (\ref{eq:metric}) but now the functions $m$ and $S$ depend on both $t$ and $r$.
With this choice of gauge potential (\ref{eq:pertpot}), the perturbation equations decouple into two sectors:
\begin{itemize}
\item
the {\em {sphaleronic sector}} consisting of entries of ${\cal {B}}$ and the functions $\gamma _{j}$;
\item
the {\em {gravitational sector}} which consists of the perturbations of the metric functions $\delta m$ and
$\delta S$ and the perturbations of the gauge field functions $\delta \omega _{j}$.
\end{itemize}
The form of the perturbation equations in the sphaleronic sector is little changed from the
asymptotically flat case \cite{brodbeck}.
It consists of $2N-1$ coupled equations for the $2N-1$ variables ($N$ diagonal entries of
the matrix ${\cal {B}}$ and $N-1$ functions $\gamma _{j}$).
In addition, there is the {\em {Gauss constraint}}, which gives $N$ coupled consistency conditions.
After much algebra (along the lines of \cite{brodbeck}), the sphaleronic sector perturbation equations
can be cast in the form
\begin{equation}
-{\ddot {{\bmath {\Psi }}}} = {\cal {U}}{\bmath {\Psi }},
\end{equation}
where a dot denotes $\partial / \partial t$,
the $(2N-1)$-dimensional vector ${\bmath {\Psi }}$ consists of combinations of perturbations
and ${\cal {U}}$ is a self-adjoint, second order, differential operator (involving derivatives with respect to $r$
but not $t$), depending on the equilibrium functions $\omega _{j}(r)$, $m(r)$ and $S(r)$.
It can be shown that the operator ${\cal {U}}$ is regular and positive provided the
unperturbed gauge functions $\omega _{j}(r)$ have no zeros and satisfy the $N-1$ inequalities
\begin{equation}
\omega _{j}^{2} >
1 + \frac {1}{2} \left( \omega _{j+1}^{2} + \omega _{j-1}^{2} \right)
\label{eq:stabineq}
\end{equation}
for all $j=1,\ldots N-1$.
These inequalities define a non-empty subset of the parameter space, which is shown in the ${\mathfrak {su}}(3)$
case in figure \ref{fig:su3stab}.

\begin{figure}[h]
\begin{center}
\includegraphics[width=6.5cm,angle=270]{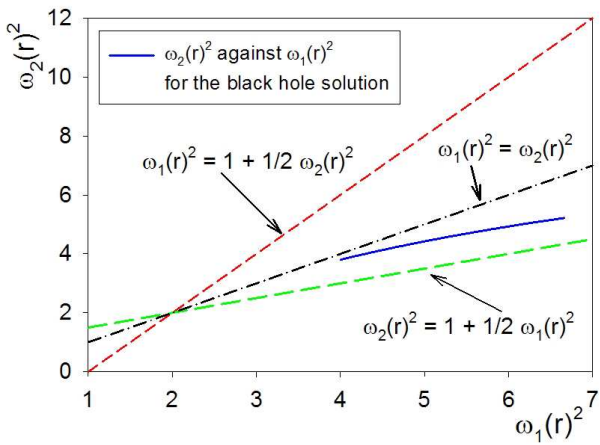}
\end{center}
\caption{An example of an ${\mathfrak {su}}(3)$ solution for which the inequalities (\ref{eq:stabineq})
are satisfied for all $r\ge r_{h}$.
In this example, $\Lambda = -10$, $r_{h}=1$ and the values of the gauge field functions at the event horizon
are $\omega _{1} ( r_{h} ) = 2$, $\omega _{2} ( r_{h} ) = 1.95$.}
\label{fig:su3stabex}
\end{figure}

The shaded region in figure \ref{fig:su3stab} shows where the inequalities (\ref{eq:stabineq}) are satisfied
for the gauge field functions at the event horizon.
However, the requirements of (\ref{eq:stabineq}) are considerably stronger, as the inequalities have to
be satisfied for {\em {all}} $r\ge r_{h}$.
Our analytic work shows that, in fact, for sufficiently large $\left| \Lambda \right| $, there
do exist solutions to the field equations for which the inequalities (\ref{eq:stabineq}) are indeed
satisfied for all $r$ (an example of such a solution is shown in figure \ref{fig:su3stabex}).
This involves proving that for at least some solutions for which the gauge field function values at the
event horizon lie within the region where the inequalities (\ref{eq:stabineq}) are satisfied,
the gauge field functions remain within this open region.

For the gravitational sector, the metric perturbations can be eliminated to yield a set of $N-1$, coupled
perturbation equations of the form
\begin{equation}
-\delta {\ddot {{\bmath {\omega }}}} = {\cal {M}}\, \delta {\bmath {\omega }},
\end{equation}
where
$\delta {\bmath {\omega }} = \left( \delta \omega _{1}, \ldots , \delta \omega _{N-1} \right) ^{T} $,
and ${\cal {M}}$ is a self-adjoint, second order, differential operator (involving derivatives with respect to $r$
but not $t$), depending on the equilibrium functions $\omega _{j}(r)$, $m(r)$ and $S(r)$.
The operator ${\cal {M}}$ is more difficult to analyze than the operator ${\cal {U}}$.
For sufficiently large $\left| \Lambda \right| $, it can be shown that ${\cal {M}}$ is a positive operator
for embedded ${\mathfrak {su}}(2)$ solutions, provided that $\omega ^{2}(r) > 1$ for all $r$
(the existence of such ${\mathfrak {su}}(2)$ solutions is proved, for sufficiently large $\left| \Lambda \right| $,
in \cite{ew}).
As described above, our analytic work ensures the existence of genuinely ${\mathfrak {su}}(N)$
solutions in a sufficiently small neighborhood of these embedded ${\mathfrak {su}}(2)$ solutions.
These ${\mathfrak {su}}(N)$ solutions are such that the inequalities (\ref{eq:stabineq}) are satisfied for
all $r\ge r_{h}$ (and therefore the solutions are stable under sphaleronic perturbations).
The positivity of ${\cal {M}}$ can then be extended to these genuinely ${\mathfrak {su}}(N)$ solutions
using an analyticity argument, based on the nodal theorem of \cite{amann}.
The technical details of this argument will be presented elsewhere.
Therefore at least some of our solutions are linearly stable in both the gravitational and sphaleronic
perturbation sectors.

For sufficiently large $\left| \Lambda \right| $ (for each fixed $r_{h}$),
we have shown the existence of ${\mathfrak {su}}(N)$
EYM black holes in adS, which are described by $N+1$ parameters and are stable under linear, spherically symmetric
perturbations.
If the cosmological constant is very large and negative, there are potentially a very large number of possible
gauge field configurations giving the same mass and magnetic charge at infinity.
As explained in the introduction,
we anticipate that these solutions may well have interesting consequences for the adS/CFT correspondence \cite{adsCFT}.
We hope to return to these questions in the near future.

\begin{acknowledgments}
We thank Eugen Radu for many informative discussions. The work of JEB is supported by UK EPSRC, and
the work of EW is supported by UK PPARC, grant reference number PPA/G/S/2003/00082.
\end{acknowledgments}

\end{document}